\documentclass[a4paper]{spie}  
\usepackage[]{graphicx}
\usepackage[]{amsmath}

\title{Coupling light into optical fibres near the diffraction limit} 

\author{Anthony J. Horton and Joss Bland-Hawthorn
\skiplinehalf
Anglo-Australian Observatory, PO Box 296, Epping, NSW, Australia
}

\authorinfo{Further author information: (Send correspondence to A.J.H.)\\A.J.H.: E-mail: ajh@aao.gov.au, Telephone: +61 2 9372 4892\\  J.B-H.: E-mail: jbh@aao.gov.au, Telephone: +61 2 9372 4851}

\begin{document} 
  \maketitle 

\begin{abstract}
The burgeoning field of astrophotonics explores the interface between 
astronomy and photonics. Important applications include photonic OH 
suppression at near-infrared wavelengths, and integrated photonic 
spectroscopy.  These new photonic mechanisms are not well matched to 
conventional multi-mode fibre bundles, and are best fed with single or
few-mode fibres.
We envisage the largest gains in astrophotonics will come from instruments 
that operate with single or few mode fibres in the diffraction limited or 
near diffraction limited regime. While astronomical instruments have largely 
solved the problem of coupling light into multi-mode fibres, this is largely 
unexplored territory for few-mode and single-mode fibres.  Here we 
describe a project to explore this topic in detail, and present initial 
results on coupling light into single and few-mode fibres at the diffraction
limit.  We find that fibres with as few as $\sim5$ guided modes have 
qualitatively different behaviour to single-mode fibres and share a number
of the beneficial characteristics of multi-mode fibres.
\end{abstract}


\keywords{few-mode, fibres, coupling, diffraction limit, astrophotonics, optical and infrared astronomy}

\section{INTRODUCTION}
\label{sect:intro}  

Optical fibres have been in use in astronomical instrumentation for almost 
30 years. They were first used for fibre-fed multi-object spectroscopy, which 
began with the Medusa instrument at Steward Observatory in 
1979\cite{Hilletal80}.  
The efficiency gains from 
the ability to observe large numbers of objects at once have made many 
otherwise impractical scientific programmes possible, including several huge 
spectroscopic surveys of great importance, for example the 2dF 
galaxy\cite{2dFGRS01} \& QSO\cite{2QZ04} redshift surveys and the Sloan 
Digital Sky Survey spectroscopy component\cite{SDSS00}. Another important 
application is integral field spectroscopy.
The first instruments using fibre-based integral field units were 
constructed in the late 
1980s\cite{MF40,Silfid,DensePak}, and this
is now a well established 
technique 
employed by a number 
of
major optical and near infrared instruments (e.g.\ GMOS\cite{GMOSIFU}, 
VIMOS\cite{VIMOS}, IMACS\cite{IMACS}, CIRPASS\cite{CIRPASS}).
Optical fibres are also in use in astronomical optical interferometers
\cite{ShaklanRoddier88}.

The importance of optical fibres in astronomy is set to increase 
even further with the continuing development of adaptive optics (AO).  
AO systems are now in place on all the world's largest astronomical telescopes,
and advanced multi-conjugate adaptive optics (MCAO) systems are planned
for both the current generation of telescopes and all of the proposed 
Extremely Large Telescopes (ELTs).  These MCAO systems will
provide large AO-corrected fields of view, and the very large number of 
spatially resolved elements will make efficient sampling of the focal plane 
essential\cite{BlandHawthorn06}.  This will require new developments in 
multi-object spectroscopy 
systems, deployable integral field units and deployable imaging systems, and
optical fibres are well suited to play an important role in all of these.  
Consequently optical fibres, and in particular optical 
fibres combined with adaptive optics systems, are extremely important to the
future of astronomy.

The majority of fibre-fed instruments so far have been designed for operation 
under natural seeing conditions, and the relative ease of coupling light into 
multi-mode fibres (MMFs), especially in the presence of 
atmospheric aberrations, has led to their exclusive use in instruments to 
date.  Developments in astrophotonics now provide a strong motivation to 
move away from MMFs, however.  Astrophotonics is a broad term used 
for the astronomical application of a wide range of photonic 
technology, and this 
burgeoning field has the potential to revolutionise 
astronomy.  Important examples which is likely to have a significant
impact in the near future are integrated photonic 
spectrographs\cite{OtherPaper} and OH suppression fibres based on aperiodic 
fibre Bragg 
gratings\cite{OHsuppr} (AFBGs), which promise to give near infrared 
instruments sky backgrounds as low or possibly even lower than those seen in 
the optical. 
These new devices are capable of 
greatly benefitting astronomy, however a significant obstacle to realising 
this potential is the fact that many have been conceived as single-mode 
devices, meaning that they cannot be fed by conventional MMFs.

The most obvious
way to integrate a single-mode photonic device into an instrument is to feed
it with a single-mode fibre (SMF), but this approach has its own difficulties.
Shaklan \& Roddier\cite{ShaklanRoddier88} and 
Coud\'e du Foresto et al\cite{CoudeduForesto00} have shown that the 
theoretical 
maximum efficiency with which a stellar image can be coupled into a single mode
fibre is $\sim 80\%$ in the absence of any atmospheric turbulence 
effects or obstructions in the telescope pupil.  When the effect of a 
reasonable circular central obstruction of 20\% of the primary diameter is included 
this falls to $\sim 70\%$, and the presence of atmospheric aberrations further
reduces the coupling efficiency in proportion to the Strehl ratio.  As a result
direct coupling of large
telescopes 
to SMFs in natural seeing is rendered impractical by low 
efficiencies, while the use of SMFs with AO places strong constraints
on the necessary performance of the AO system, both in terms of the average 
Strehl achieved and its variability (which impacts on calibration).

While the continuing development of AO may
allow highly efficient and stable coupling of telescopes to SMFs in the future
there is an intermediate approach which should allow the efficient integration
of astrophotonic devices now.  
Though many important devices are 
single-mode it is in general possible to extend them to 
operate with a few propagating modes.  With OH suppressing fibres, for example,
the atmospheric emission lines must be blocked separately for each propagating 
mode, which can be achieved either with a single, more complex AFBG or by
using converters to connect to multiple single-mode AFBGs\cite{Converter}.  
An integrated spectrograph can also be made
to work with a few modes\cite{OtherPaper}, at least at low and moderate 
resolutions.  Using modified astrophotonic devices such as these makes it is 
possible to use few mode fibres (FMFs) instead of SMFs.  The coupling of light 
into FMFs is relatively unexplored territory, however as the number of modes 
increases it will become easier to couple light into the fibres, 
which is expected to reduce the sensitivity to 
AO system performance at the expense of increasing the required complexity
of attached astrophotonic devices.  Indeed for some, such as
OH suppression fibres, it will be practical to use sufficiently 
many modes to allow use under natural seeing.  In this paper we describe an 
ongoing investigation into the trade off
between coupling efficiency and the number of propagating modes.

\section{THEORY}
\label{sect:theory}

\subsection{Fibre Modes}

The multi-mode fibres typically used in astronomical instruments, whose core
radii are much greater than the operating wavelength, can in many ways be 
regarded as `light pipes', and their properties predicted with reasonable 
accuracy using geometric optics methods.  If the size of the fibre core is 
less than $\sim 50$ times the wavelength, however, the 
geometric optics approximation breaks down and it becomes necessary 
to treat the fibre as
a dielectric waveguide and solve for the electromagnetic fields of the 
propagating modes.  The behaviour of the fibre can then be predicted from the 
characteristics of the individual modes.  In this investigation we have 
followed the treatments of Gloge\cite{Gloge71}, Midwinter\cite{Midwinter} and
Jeunhomme\cite{Jeunhomme83}.

We consider a conventional silica step-index fibre, 
consisting of a core of radius $a$ with uniform refractive index 
$n_1$ surrounding by cladding material of uniform index $n_2 < n_1$.  
In the limit of $\Delta = (n_1 - n_2) / n_2 \ll 1$ the propagating modes of 
such a fibre have a particularly simple 
form.  In this `weakly-guiding' limit the Maxwell equations can be transformed 
into a scalar 
wave equation for the longitudinal components, and the fields within the fibre
expressed as a series of linearly polarised (LP) modes.  In practice this 
is a 
reasonable approximation for a real fibre, as the difference in refractive 
indices is generally $<1\%$ and the resulting error in mode characteristics 
$<0.1\%$.

The LP modes are characterised by two numbers, the azimuthal order, $l$, and 
the radial order, $m$.  The transverse component of the electric field of the 
LP$_{lm}$ mode is given by
\begin{eqnarray}
  \label{eqn:core}
  E_{lm}(\rho,\theta) = & 
  A_{lm}\left(\sin l\theta,\cos l\theta \right)J_l(u_{lm}\rho)/J_l(u_{lm})
  & \rho \le 1 \\
  \label{eqn:cladding}
  & 
  A_{lm}\left(\sin l\theta,\cos l\theta \right)K_l(w_{lm}\rho)/K_l(w_{lm})
  & \rho > 1,
\end{eqnarray}
where $\rho$ is the normalised radial coordinate $r/a$, $J_{l}$ is the Bessel 
function of the first kind of order $l$ and $K_{l}$
is corresponding modified Bessel function of the second kind.  
For each LP mode 
with $l \ne 0$ two independent orientations exist, 
with $\sin l\theta $ and $\cos l\theta$ azimuthal dependences.  
The longitudinal components
are small compared to the transverse (by a factor $>1/\sqrt{2\Delta}$) and
can for most purposes be neglected.

The transverse propagation constants for the core ($u_{lm}$) and 
cladding ($w_{lm}$) are determined by the normalised frequency, $V$, which   
is defined by $V = 2 \pi a {\rm NA} / \lambda$ where $\lambda$ is wavelength 
and $\rm{NA} = \sqrt{n_1^2 - n_2^2} \approx n_2\sqrt{2\Delta}$ is the numerical aperture of the fibre.
The transverse propagation constants satisfy
\begin{equation}
\label{eqn:vuw}
V = \sqrt{u^2 + w^2}
\end{equation}
and $u_{lm}$,$w_{lm}$ are given by the $m$th root of
\begin{equation}
  \label{eqn:tpc}
  u\frac{J_{l-1}(u)}{J_{l}(u)} + w\frac{K_{l-1}(w)}{K_{l}(w)} = 0.
\end{equation}
From Eqs.~(\ref{eqn:vuw}) and (\ref{eqn:tpc}) it can be shown that $u_{lm}$ 
must lie between the 
$m$th zero of $J_{l-1}$ and the $m$th zero of $J_l$.  For a mode to be guided
by the fibre 
$w_{lm}$ must be real, and so the minimum value of
$u_{lm}$ defines the cutoff frequency, $V_c$, for the mode.  In the special 
case of $l=0$ the $m=1$ mode has a cutoff frequency of zero, this is the 
fundamental mode of the fibre and is always present.  For a more rigorous 
treatment of the cutoff conditions see, for example, Marcuse\cite{Marcuse}.
The cutoff frequencies
of the first few LP modes are given in Table \ref{tab:cutoffs}.
As can be seen from the table the cutoff frequencies of the LP
modes become more closely spaced at higher frequencies, in fact the number of
guided modes at a normalised frequency $V$ is approximately proportional
to $V^2$.  
\label{sect:fibremodes}
\begin{table}
\caption{Cutoff frequencies for the linearly polarised modes of a step index fibre.} 
\label{tab:cutoffs}
\begin{center}       
\begin{tabular}{|ccr|ccr|ccr|ccr|} 
\hline
\rule[-1ex]{0pt}{3.5ex} $l$ & $m$ & $V_c$ & $l$ & $m$ & $V_c$ & $l$ & $m$ & $V_c$ & $l$ & $m$ & $V_c$ \\
\hline
\rule[-1ex]{0pt}{3.5ex} 0 & 1 & \ 0.000 & 4 & 1 & \ 6.380 & 6 & 1 & \ 8.771 & 5 & 2 & 11.065 \\
\rule[-1ex]{0pt}{3.5ex} 1 & 1 & \ 2.405 & 0 & 3 & \ 7.016 & 4 & 2 & \ 9.761 & 8 & 1 & 11.619 \\
\rule[-1ex]{0pt}{3.5ex} 0 & 2 & \ 3.832 & 2 & 2 & \ 7.016 & 7 & 1 & \ 9.936 & 1 & 4 & 11.792 \\
\rule[-1ex]{0pt}{3.5ex} 2 & 1 & \ 3.832 & 5 & 1 & \ 7.588 & 0 & 4 & 10.173 & 9 & 1 & 12.225 \\
\rule[-1ex]{0pt}{3.5ex} 3 & 1 & \ 5.136 & 3 & 2 & \ 8.417 & 2 & 3 & 10.173 & 6 & 2 & 12.339 \\
\rule[-1ex]{0pt}{3.5ex} 1 & 2 & \ 5.520 & 1 & 3 & \ 8.654 & 3 & 3 & 11.086 & 4 & 3 & 13.015 \\
\hline
\end{tabular}
\end{center}
\end{table} 
\subsection{Image Field}
\label{sect:diff}

The field distribution in the focal plane of a telescope is related to the 
distribution in the entrance pupil by
\begin{equation}
  \boldsymbol{E}_{\rm focus}(\boldsymbol{r}) \propto 
  \boldsymbol{\tilde E}_{\rm pupil}\left(
  \frac{\boldsymbol{r}}{\lambda f}\right)
\end{equation}
where the tilde represents a Fourier transform and $f$ is the effective focal
length of the telescope.  The field in the entrance pupil can be written as
\begin{equation}
  \boldsymbol{E}_{\rm pupil}(\boldsymbol{r'}) = 
  \boldsymbol{E}_\star P(\boldsymbol{r'}) \boldsymbol{\Psi}(\boldsymbol{r'})
\end{equation}
where $\boldsymbol{E}_\star$ is the field received from the source, $P$ is the
pupil transmission function and $\boldsymbol{\Psi}$ is a phase screen 
which incorporates any optical aberrations, either fixed aberrations inherent 
to the telescope optics or random, variable aberrations due to passage of the 
incoming light through turbulence in the atmosphere.

Initially we consider the diffraction limited case where there are no 
aberrations, i.e.\ $\boldsymbol{\Psi} = 1$.   If we represent the telescope
pupil as a circular aperture of radius $R$ with a central obstruction of 
radius $\alpha R$ then the focal plane field distribution for a point source
is given by
\begin{equation}
  \label{eqn:diff}
  \boldsymbol{E}_{\rm focus} = \boldsymbol{E}_0 \left[ 
    \frac{2J_1(s)}{s} - \alpha^2 \frac{2J_1(\alpha s)}{\alpha s}, 
    \right]
\end{equation}
where $s$ is a scaled radial coordinate given by 
$s = 2\pi Rr/\lambda f$.  Noting that $2R/f$ is the definition of focal
ratio of the telescope, $F$, we see that the dimensions of the diffraction
limited image scale in proportion to the product $\lambda F$.

\subsection{Coupling Efficiency}

The fraction of incident energy which is coupled into an individual
mode, $\epsilon_{lm}$, can be calculated from the electric field distribution
of the image $E_{\rm focus}$ and that of the fibre mode $E_{lm}$ according to
\begin{equation}
  \label{eqn:couple}
  \epsilon_{lm} = \frac
	   {\left|\int_\infty{
	       \boldsymbol{E}^*_{\rm focus}.\boldsymbol{E}_{lm}{\rm d}A
	     }\right|^2}
	   {
	     \int_{\infty}{\left|\boldsymbol{E}_{\rm focus}\right|^2{\rm d}A} 
	     \int_{\infty}{\left|\boldsymbol{E}_{lm}\right|^2{\rm d}A}
	   }.
\end{equation}
The asterisk indicates complex conjugation and the integrals are performed over
an infinite plane at the telescope focus.  When ignoring polarisation and by
noting that the electric field of the incident light is purely transverse we
can rewrite Eq.~(\ref{eqn:couple}) in terms of inner products of 
the transverse field component distributions, i.e.
\begin{equation}
  \label{eqn:inners}
  \epsilon_{lm} = \frac
	  {\left| \left< E_{\rm focus} | E_{lm} \right> \right|^2}
	  {\left<E_{\rm focus}|E_{\rm focus}\right>
	    \left< E_{lm}|E_{lm}\right>}
\end{equation}
where $E_{\rm focus}$ is the scalar part of Eq.~\ref{eqn:diff} and 
$E_{lm}$ is the
transverse component of the LP$_{lm}$ mode field as in Eqs.~(\ref{eqn:core}) 
and (\ref{eqn:cladding}).  
In writing Eq.~(\ref{eqn:inners}) we have also used the 
fact that the longitudinal field components of an LP mode are small relative
to the transverse components (see Sect.~\ref{sect:fibremodes}) to neglect their
contribution to $\int_{\infty}{\left|\boldsymbol{E}_{lm}\right|^2{\rm d}A}$.

The coupling efficiency $\epsilon$, defined as the fraction of the incident 
energy which ends up being guided by the fibre, is simply given by summing the
$\epsilon_{lm}$ over all the guided modes.

\section{INITIAL RESULTS}
\label{sect:results}

\subsection{Model System}

In the calculations described here we use diffraction limited images 
coupled directly into a step index fibre in the focal plane.  Such a system 
can be characterised by three dimensionless parameters, 
$S$, $\alpha$ and $V$.  The parameter $S=\lambda F/a$ determines the scale of 
the image relative to the fibre core, the central obstruction size $\alpha$ 
determines the form of the 
image, and the normalised frequency $V=2\pi a\rm{NA}/\lambda$ determines the 
number and form of the guided modes.  However instead of presenting our 
results in terms of these generalised parameters we prefer to work
with the more physical set of variables $F$, $\alpha$ and 
$d=2a$, and assume appropriate values for $\lambda$ and NA.  The results can
of course be transformed to other wavelengths or NAs via the corresponding 
values of $S$ and $V$.

The wavelengths and NA used have been chosen to be representative of
likely values for an astrophotonics application.
The H-band wavelength region is of particular interest as this is the region 
being targeted by the first generation of OH suppression AFBGs, and FMF 
coupling will play an important role in integrating these devices into full 
scale 
astronomical instruments.  For this reason we have used H-band wavelengths 
in these calculations, selecting 1.5$\mu$m as representative.
Ordinary silica optical fibres typically have numerical apertures in the range 
0.1--0.2, with SMFs generally towards the lower end of this range and MMFs 
higher.  For these calculations we have used an NA of 0.1. 

\subsection{Maximum Coupling Efficiency Versus Core Diameter}
   \begin{figure}
   \begin{center}
   \begin{tabular}{c}
   \includegraphics[width=\textwidth]{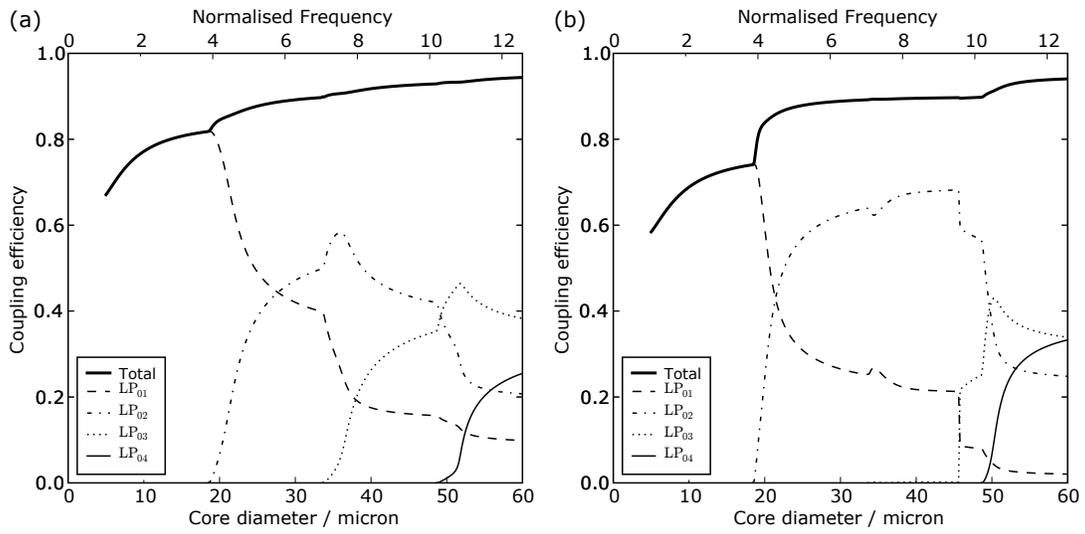}
   \end{tabular}
   \end{center}
   \caption[effvsd]
   { \label{fig:effvsd}
Maximum coupling efficiency versus core diameter for an $\rm{NA}=0.1$ fibre at
a wavelength of 1.5$\mu$m.  The overall coupling efficiency and the 
contributions from each fibre mode are shown for (a) $\alpha=0$ and 
(b) $\alpha=0.2$.}
   \end{figure}
   \begin{figure}
   \begin{center}
   \begin{tabular}{c}
   \includegraphics[width=\textwidth]{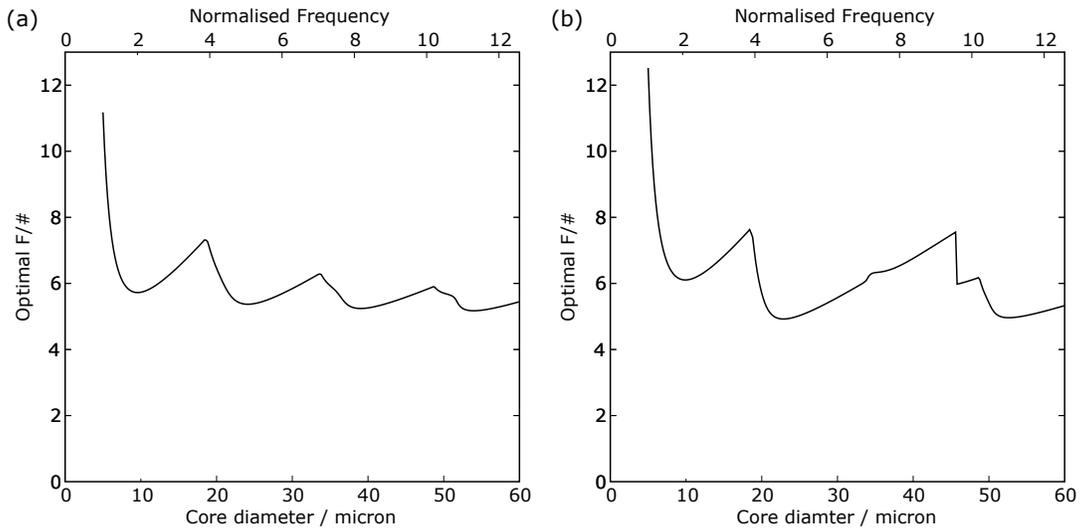}
   \end{tabular}
   \end{center}
   \caption[optFvsd]
   { \label{fig:optFvsd}
Optimal focal ratio versus core diameter for an $\rm{NA}=0.1$ fibre at 
a wavelength of 1.5$\mu$m.  The focal ratio corresponding to maximum 
coupling efficiency is shown for (a) $\alpha=0$ and (b) $\alpha=0.2$.}
   \end{figure}
   \begin{figure}
   \begin{center}
   \begin{tabular}{c}
   \includegraphics[width=\textwidth]{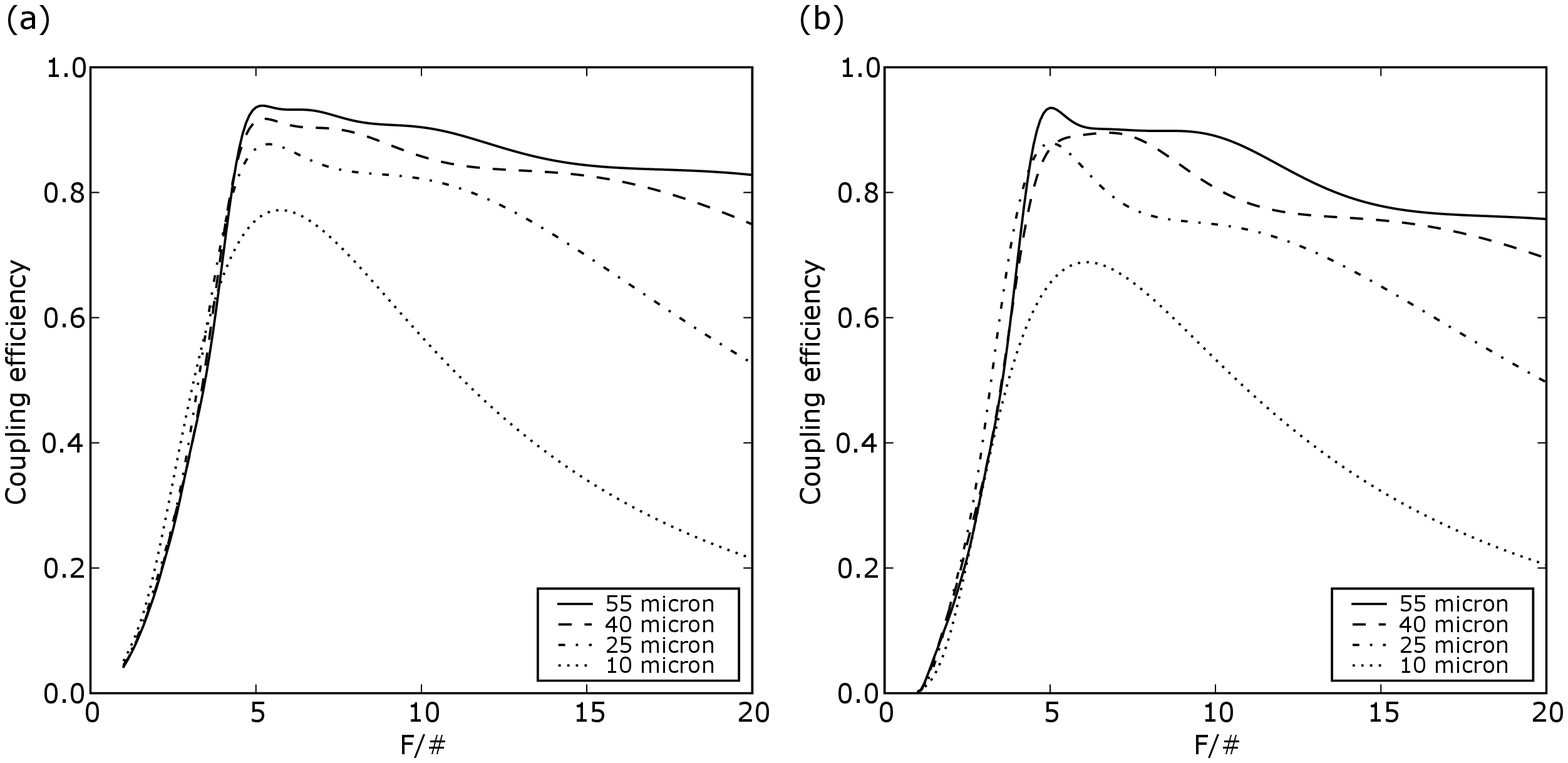}
   \end{tabular}
   \end{center}
   \caption[effvsF]
   { \label{fig:effvsF}
Coupling efficiency versus focal ratio for an $\rm{NA}=0.1$ fibre at a
wavelength of 1.5$\mu$m.  Coupling efficiency is shown for core diameters
of 10, 25, 40 and 55$\mu$m with (a) $\alpha=0$ and (b) $\alpha=0.2$.}
   \end{figure}
The first result from these investigations was the dependence of 
maximum 
coupling efficiency on the fibre core diameter.  To calculate this the 
coupling efficiency was optimised against focal ratio for core diameters in 
the range 5--60$\mu$m.  For $\lambda=1.5\mu$m and $\rm{NA}=0.1$ this range of
core diameters corresponds to a range in normalised frequency of 1.05--12.6,
which as can be seen from Table~\ref{tab:cutoffs} covers single mode operation
at the lower end ($d<11.5\mu$m) up to 23 guided modes at the upper limit.

The results are shown in Fig.~\ref{fig:effvsd}(a) and 
Fig.~\ref{fig:effvsd}(b) for $\alpha=0$ and $\alpha=0.2$ respectively.  
The two values of $\alpha$ represent the ideal case and one more
typical of the current generation of large telescopes.
In both cases there is a 
monotonic increase in total coupling efficiency as the core diameter increases
(from 67\% to 94\% for $\alpha=0$, 58\% to 94\% for $\alpha=0.2$) 
punctuated by a number of discontinuities in the gradient.  
From these results we immediately see that FMFs give significantly higher
maximum coupling efficiency than SMFs ($d<11.5\mu$m), especially in the more realistic
$\alpha=0.2$ case.
The cause of the 
discontinuities becomes apparent when the contributions from each of the 
guided modes are 
examined so these have also been plotted on Fig.~\ref{fig:effvsd}.
In this calculation the image is aligned with
the fibre axis and due to the axial symmetry only $l=0$ 
modes contribute.  In the  $\alpha=0$ case the position of the 
discontinuities coincides with the cutoff frequencies of the LP$_{02}$, 
LP$_{03}$ and LP$_{04}$ modes, and it is the appearance of these modes 
that provides a boost to the coupling efficiency.  The $\alpha=0.2$ case is 
slightly more complex as the 
different image profile results in significant jumps in coupling 
efficiency at the appearance of the LP$_{02}$ and LP$_{04}$ modes while the 
LP$_{03}$ mode does not contribute until well above its cutoff.

The significant negative effect on SMF coupling efficiency 
of a central obstruction in the telescope pupil was previously noted by Coude 
du Foresto 
et al\cite{CoudeduForesto00}, and a comparison of the small core diameter ends
of Fig.~\ref{fig:effvsd}(a) and Fig.~\ref{fig:effvsd}(b) shows that an 
efficiency loss of $\sim 10\%$ occurs with $\alpha=0.2$.  The effect of the
central obstruction on the image is to reduce the field strength in the 
central peak while increasing it in the first ring, which reduces the coupling
to the LP$_{01}$ mode.  This change in image profile increases the coupling to
the LP$_{02}$ mode, however, and so above the LP$_{02}$ cutoff we see the
maximum coupling efficiency for $\alpha=0.2$ approximately equal that for
$\alpha=0$.  Similarly the $\alpha=0.2$ image does not couple well to the 
LP$_{03}$ mode, but does to LP$_{04}$.

\subsection{Focal Ratio Dependence}

In calculating the maximum coupling efficiency as a function of core diameter
we also obtained the focal ratios at which maximum coupling efficiency
is achieved.  These optimal focal ratios are shown in 
Figs.~\ref{fig:optFvsd}(a) and \ref{fig:optFvsd}(b) for the $\alpha=0$ and
$\alpha=0.2$ cases.  To get a fuller picture of the effect of focal ratio
we also calculated coupling efficiency for $F=1$--20
for four core diameters of 10, 25, 40, 55$\mu$m.  With
$\rm{NA}=0.1$ and $\lambda=1.5\mu$m these core diameters
have one, two, three and four guided $l=0$ modes respectively.  The results
for $\alpha=0$ and $\alpha=0.2$ are plotted as Fig.~\ref{fig:effvsF}(a) and
Fig.~\ref{fig:effvsF}(b).  

The mode cutoffs and changes in the contributions of each mode seen in 
Fig.~\ref{fig:effvsd} are also seen reflected in variations in the 
optimal focal ratio in Fig.~\ref{fig:optFvsd}, however as the diameter 
increases the variations diminish and the optimal focal ratio converges 
from above on a value of $\sim 5$.  In other words the optimal image size does
not scale with the fibre diameter but instead remains relatively constant.
Looking at Fig.~\ref{fig:effvsF} we see 
that for all 
the plotted core diameters the coupling efficiency declines rapidly for focal 
ratios less than $\sim5$.  
This low $F$ cutoff is the same as predicted in the multi-mode limit via 
geometric optics.  By considering the condition for total 
internal reflection to occur at the core/cladding interface the
numerical aperture of the fibre can be 
equated to the sine of the maximum angle of incidence at which a light ray can 
strike the fibre face and still be guided within the core.  
This maximum angle can then be converted into a minimum focal ratio via 
$F_{\rm min} = 1/2\tan\left(\sin^{-1}\rm{NA}\right)\approx 1/(2.\rm{NA})$, 
and for an NA of 0.1 this is 4.97.
So we see that the geometric optics derived maximum focal ratio for MMFs also 
applies to the single and few-mode regime, and furthermore the optimum focal
ratio for FMFs lies close to this value.

The other main feature of Fig.~\ref{fig:effvsF} is the decreasing 
sensitivity of the coupling efficiency to higher than optimum focal ratios as
the core diameter increases.  This would be intuitively expected as the larger 
core allows for greater magnification of the image (higher $F$) before a 
significant proportion of the light falls outside the core.  Upon examining the
contributions of the individual modes as the focal ratio changes it is also 
possible to explain a number of the features in the coupling efficiency slope,
which are particularly apparent in the $\alpha=0.2$ case.  As the focal ratio
increases the lower order modes become progressively more important and so, 
for 
instance, the significant decline in coupling efficiency between $F=5$ and 
$F=7$ 
for a 25$\mu$m core when $\alpha=0.2$ is due to a decline in the LP$_{02}$ 
mode and a transition to dominance by the poorly coupling LP$_{01}$ mode.  The 
other features can be similarly ascribed to the decline of particular modes.

\subsection{Sensitivity to Decentring}
   \begin{figure}
   \begin{center}
   \begin{tabular}{c}
   \includegraphics[width=0.625\textwidth]{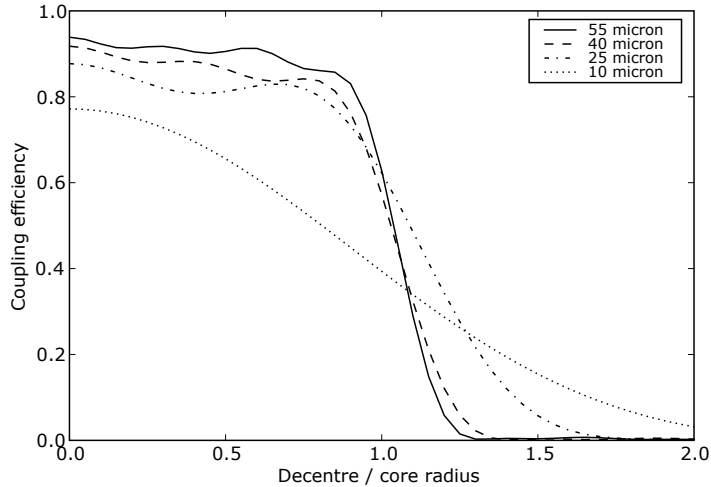}
   \end{tabular}
   \end{center}
   \caption[effvsC]
   { \label{fig:effvsC}
Coupling efficiency versus decentring of the image centre from the fibre axis.
  Coupling efficiency is shown for $\rm{NA}=0.1$ fibres with 10, 25, 40 and 
55$\mu$m core diameters at a wavelength of 1.5$\mu$m and with $\alpha=0$.}
   \end{figure}
While the increases seen in maximum coupling efficiency with only a few 
additional modes are pleasing, the primary motivation for undertaking this 
investigation is the expectation that FMF coupling efficiency will be
considerably less sensitive than SMFs to deviations from ideal conditions,  
thereby making them a more practical option for real world applications.  
In order to
determine to what extent this is true we must calculate the effect of
realistic aberrations on the coupling efficiency. However, at the time
of writing, the only aspect of this analysis which is complete is an
examination of the effect of displacing the image centre from the fibre axis. 
Such displacements are certain to occur to some degree in astronomical use 
for a number of reasons such as imperfect correction of wavefront tip/tilt, 
finite fibre positioner precision and finite accuracy of astrometry.

Figure~\ref{fig:effvsC} shows the coupling efficiency as a function of image
decentring for 10, 25, 40 and 55$\mu$m diameter cores.  
For these calculations there is no axial symmetry and so all the guided modes
must be included, not just the $l=0$ ones.
With $\rm{NA}=0.1$ and $\lambda=1.5\mu$m the four core diameters give 1, 5, 10 
and 19 guided
modes respectively.  
The previously calculated focal ratios for maximum (on-axis) coupling 
efficiency were used for each diameter.  For simplicity only the 
$\alpha=0$ case is shown, but the
$\alpha=0.2$ results are similar.  There is a qualitative difference in 
behaviour here between SMFs and FMFs, whereas fibres with 5 or more 
modes show variations in coupling efficiency of only $\sim 5\%$ for 
decentres up to 80\% of the core radius the single-mode fibre loses over a 
third of its coupling efficiency over the same range.  This flat response is a 
desirable property of the FMFs as it means that moderate image-fibre 
misalignments will not cause drops in throughput and attendant 
calibration issues.

Knowing the dependence of point source coupling efficiency on image position 
also enables
us to calculate the coupling efficiency for an extended source. 
For an extended source with an angular intensity 
distribution on the sky of $I_{\rm ext}(\boldsymbol{\theta})$ the effective
coupling efficiency $\epsilon_{\rm ext}$ is given by
\begin{equation}
  \epsilon_{\rm ext} = \frac
	  {\int_\infty{\epsilon_{\rm psf}(\boldsymbol{r})
	      I_{\rm ext}(\boldsymbol{r}/f)\rm{d}A}}
	  {\int_\infty{I_{\rm ext}(\boldsymbol{r}/f)\rm{d}A}}
\end{equation}
where $\epsilon_{\rm psf}(\boldsymbol{r})$ is the coupling efficiency for a 
point source image centred on a position $\boldsymbol{r}$ in the focal plane.  
This is in effect taking a weighted average of the coupling efficiency over
the area of the source image, and so while the exact figure will 
depend on the shape and size of the source what can be said is that the 
flatter responses seen with the FMFs will further increase their coupling 
efficiency advantage over SMFs for resolved sources.

We can also calculate the effective solid angle of sky coupled into the 
fibre, $\Omega_{\rm sky} = \pi(a/f)^2\int_\infty{\epsilon_{\rm psf}\rm{d}A}$.  
For the core diameters considered here we find
$\int_\infty{\epsilon_{\rm psf}\rm{d}A} \approx 1$ (0.98, 1.10, 0.93 and 0.96) 
and so 
$\Omega_{\rm sky} \approx \pi D^2(a/F)^2$, the same result as would be derived
from geometric optics.  This strong dependence of the sky background on 
$(a/F)^2$ together with the previously established relative insensitivity of 
FMF coupling efficiency to increasing $F$ (see Fig.~\ref{fig:effvsF}) means 
that
the focal ratio which maximises signal-to-noise ratio will in general be 
larger than the optimal value for coupling efficiency, and except for 
very bright sources will be that which matches the scale of the source 
image to the core size, the same as for the MMFs.

\subsection{Wavelength Dependence}
   \begin{figure}
   \begin{center}
   \begin{tabular}{c}
   \includegraphics[width=0.625\textwidth]{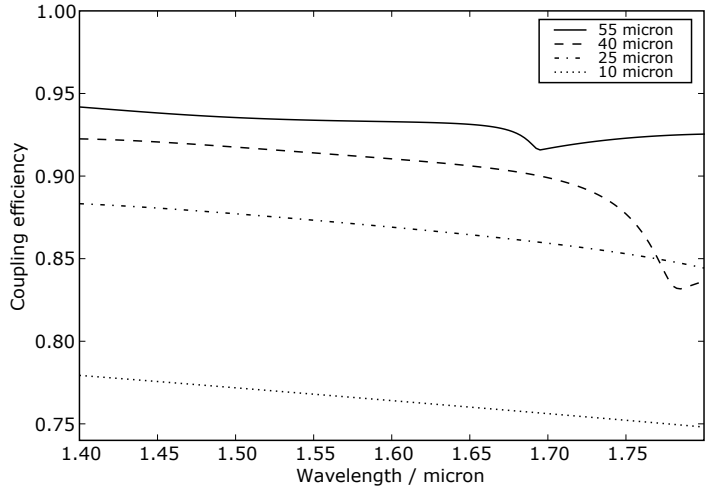}
   \end{tabular}
   \end{center}
   \caption[effvsW]
   { \label{fig:effvsW}
Coupling efficiency versus wavelength for $\rm{NA}=0.1$ fibres with 
core diameters of 10, 25, 40 and 55$\mu$m, and with $\alpha=0$.  For each 
diameter the focal ratio was fixed at the optimal value for a wavelength 
of 1.6$\mu$m.}
   \end{figure}
The previous calculations have been performed for a single wavelength, however
astronomical applications generally require good performance
over a wide wavelength range.  Figure~\ref{fig:effvsW} shows coupling 
efficiency as a function of wavelength for $\rm{NA}=0.1$ fibres with core 
diameters of 10, 25, 40 and 55$\mu$m.  For each diameter the focal ratio was 
fixed
at the optimal value for $1.6\mu$m while the wavelength was varied over the 
range 1.4--1.8$\mu$m.  This wavelength range encompasses the 
whole of the $\sim$1.45--1.75$\mu$m H-band, which is of particular interest
due to its connection with OH suppression fibre developments.
Only $\alpha=0$ 
results are shown, but $\alpha=0.2$ is similar.  The general trend for all core
diameters is a gradual decrease in coupling efficiency of $\sim3\%$ 
over the wavelength range, however for the 40 and 55$\mu$m core diameters there
is a significant dip superimposed on this steady decline.  For the 40$\mu$m 
core this feature corresponds to the LP$_{03}$ mode cutoff, while for 
55$\mu$m the LP$_{04}$ mode cutoff is the cause.  So we see that in general 
SMFs and FMFs exhibit good broadband performance, however it would be best to
chose fibre parameters so as to avoid an $l=0$ mode cutoff falling within 
an operational wavelength band.

\section{DISCUSSION}

Astrophotonic developments such as OH suppression fibres and integrated 
photonic spectrographs have enormous potential, however efficiently 
integrating them into an astronomical instrument presents a challenge.  
The MMFs conventionally used in astronomy, while efficient at accepting 
starlight, are not suitable for feeding light into these devices as the 
devices are not able to accept a large number of modes.  SMFs, on the other
hand, while ideal for feeding light into astrophotonic devices are difficult
to couple starlight into, even with adaptive optics.  We have begun an 
investigation into the intermediate territory of FMFs, in order to find the
best compromise between the two extremes.

Our initial results on diffraction limited fibre coupling have shown that 
FMFs exhibit many of the desirable properties of MMFs even when there are only 
of order 10 guided modes.  For example, FMFs offer higher maximum coupling 
efficiency than SMFs ($>90\%$), especially for extended sources.  Also FMFs 
are less sensitive to the effects of obstructions in the telescope pupil than 
SMFs are.  Unlike SMFs, FMFs can efficiently couple light over a range of 
focal ratios from $F_{\rm min} \approx 1/(2.\rm{NA})$ up to a maximum value 
determined by matching the size of the image to the fibre core.  FMFs are also 
tolerant of displacement of the image centre from the fibre axis, provided the 
image remains within the fibre core.  Both SMFs and FMFs exhibit little 
sensitivity to wavelength.

While these results are encouraging the use of perfect diffraction limited
images does represent an idealised case.  In any real ground based telescope
the adaptive optics correction will be imperfect, and the residual 
atmospheric wavefront perturbations will effect the coupling efficiency.  
It is known that SMF coupling is highly sensitive to imperfect correction, with
the coupling efficiency declining in proportion to the Strehl 
ratio\cite{CoudeduForesto00}, however the corresponding dependency for FMFs 
has not yet been investigated.  
At the time of writing simulated partially
corrected atmospheric phase screens were being included in the model system to 
investigate the effects of various levels of aberrations on fibre coupling 
performance.
The aim of this work is to determine
the dependence of FMF coupling efficiency on the order of correction, from
natural seeing to the diffraction limit, and thereby establish the number
of modes required for acceptable throughput levels under a range of realistic
usage conditions.

Preliminary results have also been obtained for pupil-plane coupling to FMFs,
which show a similar rapid convergence on MMF behaviour above $\sim10$ modes
as the image-plane results discussed here.  This work will be extended to 
model lenslet arrays of various geometries and both image and pupil-plane 
coupling with a view to determining the best approach for FMF integral field 
spectroscopy.


\acknowledgments     
The authors gratefully acknowledge the support of PPARC research grant 
PP/D002494/1 for this programme of work.


\bibliography{paper}   
\bibliographystyle{spiebib}   

\end{document}